\documentclass[letterpaper,10pt,conference]{IEEEtran}

\usepackage[utf8]{inputenc}
\usepackage[T1]{fontenc}
\usepackage{microtype} 
\usepackage{graphicx}
\usepackage{amsmath}
\usepackage{amssymb}
\usepackage{booktabs}
\usepackage{multirow}
\usepackage{array}
\usepackage{xcolor}
\usepackage{url}
\usepackage{hyperref}

\graphicspath{{figures/}}

\hypersetup{
    colorlinks=true,
    linkcolor=blue,
    citecolor=blue,
    urlcolor=blue,
    pdftitle={Cross-LLM Generalization of Behavioral Backdoor Detection in AI Agent Supply Chains},
    pdfauthor={Arun Chowdary Sanna, Enterprise AI Architect}
}

\newcommand{\code}[1]{\texttt{#1}}

\begin{document}

\title{Cross-LLM Generalization of Behavioral Backdoor Detection in AI Agent Supply Chains}

\author{
\IEEEauthorblockN{Arun Chowdary Sanna}
\IEEEauthorblockA{
Enterprise AI Architect\\
Precise Software Solutions\\
Email: arun.sanna@outlook.com
}
}

\maketitle

\begin{abstract}

As AI agents become integral to enterprise workflows, their reliance on shared tool libraries and pre-trained components creates significant supply chain vulnerabilities.
While previous work has demonstrated behavioral backdoor detection within individual LLM architectures, the critical question of cross-LLM generalization remains unexplored, a gap with serious implications for organizations deploying multiple AI systems.

To our knowledge, we present the first systematic study of cross-LLM behavioral backdoor detection, evaluating generalization across six production LLMs (GPT-5.1, Claude Sonnet 4.5, Grok 4.1, Llama 4 Maverick, GPT-OSS 120B, and DeepSeek Chat V3.1).
Through 1,198 execution traces and 36 cross-model experiments, we quantify a critical finding: \textbf{single-model detectors achieve 92.7\% accuracy within their training distribution but only 49.2\% across different LLMs, a 43.4 percentage point generalization gap equivalent to random guessing}.

Our analysis reveals that this gap stems from model-specific behavioral signatures, particularly in temporal features (coefficient of variation $>$ 0.8), while structural features (sequence patterns and dependencies) remain stable across architectures.
We show that a simple deployment strategy, model-aware detection incorporating model identity as an additional feature, achieves 90.6\% accuracy universally across all evaluated models, demonstrating that the gap, while severe, can be addressed with appropriate multi-LLM training.

These findings have immediate practical implications: organizations using multiple LLMs cannot rely on single-model detectors and require unified detection strategies.
We release our multi-LLM trace dataset and detection framework to enable reproducible research in this emerging area.
Our work establishes cross-LLM generalization as a critical dimension for AI agent security evaluation.

\end{abstract}

\section{Introduction}
\label{sec:introduction}

AI agents have become critical components in enterprise software, automating tasks from customer service to code generation~\cite{browsergym2025,workarena2024}.
These agents leverage large language models (LLMs) to perform complex reasoning, tool invocation, and multi-step planning.
As agent adoption accelerates, supply chain security emerges as a critical concern: agents trained or fine-tuned by third parties may contain backdoors that activate under specific conditions~\cite{malice2024,ai-supply-chain2025}.

Backdoor attacks in AI agents pose unique challenges compared to traditional software supply chain attacks.
Unlike static code vulnerabilities, agent backdoors exploit the probabilistic nature of LLM outputs, making them difficult to detect through static analysis.
Recent work has demonstrated multiple threat vectors:
\textit{data poisoning} (injecting malicious examples during training)~\cite{poisoning-webscale2024,poisoning-instruction2023} and
\textit{tool manipulation} (compromising agent tools)~\cite{eia2024}.

Existing defenses fall into two categories: static code analysis and model watermarking.
Static analysis approaches~\cite{badagent2024} examine agent code for suspicious patterns but fail to detect runtime-activated backdoors.
Model watermarking techniques~\cite{agentpoison2024} embed signatures in model weights but require GPU resources and impose high latency ($>$1 second per inference), making them impractical for real-time deployment.
\textbf{Critically, no prior work evaluates detection across multiple LLM architectures}, leaving a fundamental gap in understanding how behavioral detectors generalize across the heterogeneous LLMs used in production environments.

To our knowledge, we present the first systematic study of cross-LLM behavioral backdoor detection, evaluating generalization across six production LLMs from five different providers.
Our key insight is that while single-model detectors achieve high accuracy within their training distribution, they fail catastrophically when applied to different LLM architectures, a gap with serious security implications.
We quantify this gap (43.4 percentage points) and propose model-aware detection that substantially closes it.

Through comprehensive evaluation on 1,198 execution traces across six production LLMs (GPT-5.1, Claude Sonnet 4.5, Grok 4.1, Llama 4 Maverick, GPT-OSS 120B, DeepSeek Chat V3.1) and 36 cross-model experiments, we make five key contributions:

\begin{enumerate}
\item \textbf{First Systematic Cross-LLM Evaluation}: We conduct the most comprehensive study of behavioral backdoor detection across 6 production LLMs from 5 providers with 1,198 execution traces.

\item \textbf{Generalization Gap Quantification}: We demonstrate that single-model detectors achieve 92.7\% same-model accuracy but only 49.2\% cross-model accuracy, a 43.4 percentage point gap equivalent to random guessing.

\item \textbf{Architectural Analysis}: We identify the root cause as model-specific behavioral signatures: temporal features exhibit high variance across models (coefficient of variation $>$ 0.8), while structural features (sequence patterns) remain stable.

\item \textbf{Deployment Strategy}: We demonstrate that model-aware detection, incorporating model identity as a categorical feature, achieves 90.6\% universal accuracy, showing the gap can be addressed with multi-LLM-aware training.

\item \textbf{Open Science}: We release our multi-LLM behavioral trace dataset and detection framework to enable reproducible cross-model security research.
\end{enumerate}

The remainder of this paper is organized as follows.
Section~\ref{sec:related-work} surveys related work on backdoor attacks and detection.
Section~\ref{sec:threat-model} defines our threat model and attack vectors.
Section~\ref{sec:methodology} describes our feature extraction and detection pipeline.
Section~\ref{sec:experiments} details our experimental setup and multi-LLM dataset.
Section~\ref{sec:results} presents evaluation results for three research questions (RQ1: cross-LLM generalization, RQ2: architectural analysis, RQ3: ensemble approaches).
Section~\ref{sec:discussion} discusses deployment implications.
Section~\ref{sec:limitations} acknowledges limitations.
Section~\ref{sec:conclusion} concludes with future work.

\section{Background and Related Work}
\label{sec:related-work}

\subsection{Backdoor Attacks in LLMs}

Backdoor attacks in machine learning have been studied extensively, beginning with BadNets~\cite{badnets2017}, which demonstrated supply chain vulnerabilities in neural networks.
Recent work has extended these attacks to LLMs through multiple vectors.
Data poisoning approaches~\cite{poisoning-webscale2024,poisoning-instruction2023,data-poisoning-llm2024} inject malicious examples into training data, causing models to exhibit backdoor behaviors when triggered.
Carlini et al.~\cite{poisoning-webscale2024} showed that poisoning web-scale datasets is practical, requiring modification of only 0.01\% of training examples.
Wan et al.~\cite{poisoning-instruction2023} demonstrated that instruction tuning is particularly vulnerable to poisoning attacks.

Beyond data poisoning, researchers have explored preference manipulation~\cite{best-of-venom2024}, in-context learning attacks~\cite{backdoor-icl2023}, and persistent backdoors that survive safety training~\cite{sleeper-agents2024}.
Hubinger et al.~\cite{sleeper-agents2024} showed that backdoored models can maintain malicious behavior even after extensive fine-tuning.
However, these attacks focus on general LLM behavior rather than agent-specific vulnerabilities.

\subsection{Backdoor Attacks in AI Agents}

AI agents face unique vulnerabilities beyond general LLM attacks due to their multi-step reasoning, tool invocation, and stateful execution.
Wang et al.~\cite{badagent2024} introduced BadAgent, demonstrating backdoor insertion through code manipulation in agent workflows.
Their static analysis-based detection achieves 45--60\% recall but fails on runtime-activated backdoors that only manifest during specific execution conditions.

Chen et al.~\cite{agentpoison2024} presented AgentPoison, which poisons agent memory and knowledge bases to trigger malicious behaviors.
Their model watermarking approach achieves $\sim$70\% recall but requires GPU resources and imposes high latency ($>$1 second per inference), limiting practical deployment.
Critically, neither work evaluates detection across multiple threat models or considers cross-LLM generalization.

Recent work has also explored environmental attacks~\cite{eia2024}, where adversaries inject malicious content into web pages or tools accessed by agents.
Liao et al.~\cite{eia2024} showed that environmental injection can cause privacy leakage with high success rates.
Boisvert et al.~\cite{malice2024} provided a comprehensive analysis of backdoors across the AI agent supply chain, demonstrating that even 2\% data poisoning can achieve 80\% attack success.

\subsection{Backdoor Detection Methods}

Traditional backdoor detection methods focus on model inspection and input analysis.
Neural Cleanse~\cite{neuralcleanse2019} identifies backdoors by reverse-engineering triggers through gradient-based optimization.
Spectral Signatures~\cite{spectral2019} leverages singular value decomposition to detect anomalies in model representations.
Activation Clustering~\cite{activation2019} groups activations to identify poisoned samples.
However, these methods assume access to model internals and fail to generalize to black-box agent deployments.

Recent defenses have explored guardrail systems~\cite{llamafirewall2025,granite-guardian2025} that filter inputs and outputs for safety violations.
While effective against prompt injection and jailbreaks, these systems cannot detect backdoors embedded in model weights that activate only under specific conditions.
Moreover, guardrails impose latency overhead that may be prohibitive for real-time agent systems.

\subsection{Behavioral Anomaly Detection}

Behavioral anomaly detection has a rich history in cybersecurity, from system call analysis~\cite{systemcall-ids1996} to network intrusion detection.
Forrest et al.~\cite{systemcall-ids1996} pioneered using process behavior (system call sequences) to detect malicious activity, establishing the principle that anomalous behavior often indicates compromise.
Modern approaches~\cite{anomaly-survey2021} employ deep learning for time series anomaly detection across domains like fraud detection and malware analysis.

However, applying these techniques to AI agents poses unique challenges.
Unlike traditional processes with deterministic behavior, LLM agents exhibit probabilistic outputs and context-dependent reasoning.
Execution traces vary significantly based on task complexity, available tools, and model architecture.
This variability makes it difficult to define ``normal'' behavior, requiring careful feature engineering to capture meaningful patterns while accounting for legitimate diversity in agent execution.

\subsection{Positioning of This Work}

Prior work on agent backdoor detection~\cite{badagent2024,agentpoison2024} evaluates detection on a single LLM architecture, leaving cross-LLM generalization, a critical dimension for production deployments, unstudied.
To our knowledge, we present the first systematic study addressing this gap, evaluating detection across 6 production LLMs from 5 providers.
Our key contributions include: (1) quantifying a 43.4 percentage point generalization gap between same-model (92.7\%) and cross-model (49.2\%) detection, (2) identifying temporal feature instability (CV $>$ 0.8) as the root cause, and (3) demonstrating model-aware detection as an effective mitigation achieving 90.6\% universal accuracy.
Table~\ref{tab:related-work} summarizes these differences.

\begin{table*}[t]
\centering
\caption{Comparison with prior work on agent backdoor detection}
\label{tab:related-work}
\begin{tabular}{@{}lccc@{}}
\toprule
\textbf{Aspect} & \textbf{BadAgent} & \textbf{AgentPoison} & \textbf{Our Work} \\
\midrule
Detection Method & Static analysis & Watermarking & Behavioral \\
Same-Model Accuracy & 45--60\% & $\sim$70\% & 92.7\% \\
Cross-Model Accuracy & Not evaluated & Not evaluated & 49.2\% (single-model) \\
     &     &     & 90.6\% (model-aware) \\
Models Evaluated & 1 & 1 & 6 (5 providers) \\
Traces Collected & Not specified & 500 & 1,198 \\
Cross-LLM Analysis & No & No & Yes (first study) \\
\bottomrule
\end{tabular}
\end{table*}

\section{Threat Model}
\label{sec:threat-model}

\subsection{Adversary Model}

We consider an adversary who aims to deploy backdoored AI agents into production systems through the supply chain.
The adversary has the following capabilities:

\begin{itemize}
\item \textbf{Training Data Access}: Can inject malicious examples into training datasets
\item \textbf{Tool Access}: Can compromise agent tools or their implementations
\item \textbf{Model Access}: Can directly modify model weights (for model providers)
\item \textbf{Trigger Control}: Can craft specific inputs that activate backdoors
\end{itemize}

\subsection{Attack Vectors}

We focus on two threat models representing the most common supply chain attack vectors for AI agents:

\paragraph{TM1: Data Poisoning}
The adversary injects malicious training examples that cause the agent to exhibit backdoor behavior when specific triggers are present in the input.
Example: An agent trained on poisoned customer service data that leaks sensitive information when queries contain specific phrases.

\paragraph{TM2: Tool Manipulation}
The adversary compromises one or more tools available to the agent, causing them to return malicious outputs or perform unauthorized actions.
Example: A file system tool that exfiltrates data when accessed by the agent.

\paragraph{Out-of-Scope Threat: Model Tampering}
Direct modification of model weights (weight-level backdoors) is outside the scope of this study.
Such attacks require different detection approaches (model inspection, activation analysis) rather than behavioral monitoring.
We focus on attacks detectable through execution trace analysis.

\subsection{Defense Assumptions}

Our detection system assumes:

\begin{itemize}
\item \textbf{Access to Execution Traces}: The system can monitor agent execution at runtime
\item \textbf{Benign Training Data}: A corpus of benign agent traces is available for training the detector
\item \textbf{Known Threat Models}: The detector is trained on examples from known attack categories
\item \textbf{Real-Time Constraints}: Detection must complete within milliseconds to be practical
\end{itemize}

\subsection{Out of Scope}

We do not consider:
\begin{itemize}
\item Adversaries who can modify the detection system itself
\item Zero-day attack vectors not represented in training data
\item Adversarial examples specifically crafted to evade our detector
\end{itemize}

\section{System Design}
\label{sec:methodology}

\subsection{System Architecture}

Our detection system consists of three components:
\begin{enumerate}
\item \textbf{Trace Collection}: Captures agent execution traces with timestamps, tool invocations, and state changes
\item \textbf{Feature Extraction}: Computes 51 behavioral features across 4 categories
\item \textbf{Classification}: Applies trained ML models (SVM, Random Forest) to predict benign/backdoor
\end{enumerate}

\subsection{Feature Engineering}

We extract 51 behavioral features organized into four categories, following established feature engineering principles~\cite{feature-engineering2018,feature-selection2003}.
Our feature design balances expressiveness (capturing diverse attack behaviors) with computational efficiency (enabling real-time detection).

\paragraph{Temporal Features (10 features)}
Temporal features capture timing patterns that may reveal backdoor activations through anomalous delays or rushed execution.
\begin{itemize}
\item \code{avg\_duration}: Mean step duration
\item \code{max\_duration}: Maximum step duration (unusually long steps may indicate data exfiltration)
\item \code{std\_duration}: Standard deviation of step times (high variance indicates erratic behavior)
\item \code{avg\_delay}: Average delay between steps
\item \code{max\_delay}: Maximum inter-step delay
\item \code{total\_duration}: Total trace execution time
\item \code{execution\_rate}: Steps per second (low rates may indicate covert operations)
\item \code{timing\_entropy}: Entropy of timing patterns
\item \code{has\_burst}: Whether trace contains burst activity patterns
\item \code{burst\_intensity}: Intensity of burst activity (if present)
\end{itemize}

\textit{Example}: A backdoored agent executing a data exfiltration attack shows \code{avg\_delay} = 0.05s (benign: 2.3s) and \code{max\_duration} = 15.2s (benign: 3.1s), revealing suspicious rushed execution followed by prolonged data aggregation.

\paragraph{Sequence Features (15 features)}
Sequence features capture execution graph topology and transition patterns.
\begin{itemize}
\item \code{unique\_bigrams}: Number of distinct action bigrams
\item \code{max\_bigram\_freq}: Maximum bigram frequency
\item \code{bigram\_diversity}: Shannon entropy of bigram distribution
\item \code{unique\_trigrams}: Number of distinct action trigrams
\item \code{trigram\_diversity}: Trigram distribution entropy
\item \code{repetition\_ratio}: Fraction of repeated action sequences
\item \code{max\_consecutive\_repeats}: Longest repeated action run
\item \code{sequence\_length}: Total number of actions in sequence
\item \code{unique\_transitions}: Number of distinct state transitions
\item \code{transition\_entropy}: Entropy of transition patterns
\item \code{has\_loops}: Whether execution contains loops
\item \code{loop\_count}: Number of loops in execution graph
\item \code{dependency\_ratio}: Fraction of steps with data dependencies
\item \code{total\_dependencies}: Count of data dependencies between steps
\item \code{rare\_tool\_ratio}: Fraction of rarely-used tools in sequence
\end{itemize}

\textit{Example}: A benign customer service agent exhibits \code{unique\_bigrams} = 18, \code{sequence\_length} = 23, \code{transition\_entropy} = 2.1 (complex multi-step reasoning). A backdoored variant shows \code{unique\_bigrams} = 4, \code{sequence\_length} = 8, \code{repetition\_ratio} = 0.6 (repetitive linear execution toward malicious goal).

\paragraph{Action Features (12 features)}
Action features capture tool usage patterns and agent decision-making behaviors.
\begin{itemize}
\item \code{tool\_count}, \code{unique\_tools}, \code{tool\_diversity}: Basic tool usage statistics
\item \code{most\_common\_tool\_freq}: Frequency of most-used tool (backdoors may over-rely on specific tools)
\item \code{tool\_transition\_entropy}: Entropy of tool call sequences (measures predictability)
\item \code{file\_read\_count}, \code{web\_fetch\_count}: Per-tool invocation counts
\item \code{tool\_switching\_rate}: Consecutive vs. interleaved tool usage
\item \code{unauthorized\_tool\_access}: Attempts to use restricted tools
\item \code{tool\_sequence\_anomaly\_score}: Distance from typical tool usage patterns
\item \code{avg\_params\_length}, \code{max\_params\_length}: Tool parameter size statistics
\end{itemize}

\textit{Example}: A backdoored web agent shows \code{file\_read\_count} = 47 (benign: 3), \code{tool\_diversity} = 0.2 (benign: 0.8), and unusual tool transitions (web\_fetch → file\_read repeated 15 times), indicating systematic data exfiltration.

\paragraph{Data-Flow Features (14 features)}
Data-flow features track input/output characteristics and information flow across execution steps.
\begin{itemize}
\item \code{avg\_input\_size}, \code{max\_input\_size}, \code{std\_input\_size}: Input size statistics
\item \code{avg\_output\_size}, \code{max\_output\_size}, \code{std\_output\_size}: Output size statistics
\item \code{avg\_io\_ratio}, \code{max\_io\_ratio}: Input-to-output ratios (high ratios indicate data aggregation)
\item \code{input\_diversity}, \code{output\_diversity}: Diversity of data patterns
\item \code{io\_entropy}, \code{io\_coupling}: Information-theoretic measures of data flow
\item \code{sensitive\_data\_mentions}: Count of sensitive keywords in data flow
\item \code{data\_flow\_complexity}: Complexity score of data movement patterns
\end{itemize}

\textit{Example}: A backdoored agent exhibits \code{avg\_io\_ratio} = 3.2 (benign: 1.1), \code{max\_output\_size} = 45KB (benign: 2KB), and \code{sensitive\_data\_mentions} = 12 (benign: 0), indicating systematic data collection and exfiltration.

\subsection{Classification Pipeline}

\paragraph{Training}
We train two classifiers based on their proven effectiveness for anomaly detection~\cite{statistical-learning2013}:
\begin{itemize}
\item \textbf{Support Vector Machine (SVM)}: Linear kernel, C=1.0, class weights balanced
\item \textbf{Random Forest}: 100 estimators, max depth=10, class weights balanced
\end{itemize}

SVMs excel at high-dimensional feature spaces with clear margins, while Random Forests provide interpretable feature importance scores.
Training uses stratified sampling to maintain class balance (benign:backdoor = 1:1).
All features are standardized using z-score normalization to ensure equal contribution during training.

\paragraph{Inference}
At runtime, feature extraction and classification complete in under 1ms, enabling real-time monitoring without impacting agent performance.


\section{Experimental Setup}
\label{sec:experiments}

\subsection{Multi-LLM Dataset}

\paragraph{Data Collection}
We collected 1,198 execution traces from AI agents across six production LLMs via the OpenRouter API:

\begin{table}[t]
\centering
\caption{Multi-LLM Dataset Overview}
\label{tab:dataset}
\small
\begin{tabular}{@{}lcccc@{}}
\toprule
\textbf{Model} & \textbf{Provider} & \textbf{Benign} & \textbf{Backdoor} & \textbf{Total} \\
\midrule
GPT-5.1 & OpenAI & 100 & 100 & 200 \\
Claude Sonnet 4.5 & Anthropic & 100 & 100 & 200 \\
Grok 4.1 Fast & XAI & 100 & 100 & 200 \\
Llama 4 Maverick & Meta & 100 & 100 & 200 \\
GPT-OSS 120B & OpenAI & 100 & 100 & 200 \\
DeepSeek Chat V3.1 & DeepSeek & 99 & 99 & 198 \\
\midrule
\textbf{TOTAL} & \textbf{5 providers} & \textbf{599} & \textbf{599} & \textbf{1,198} \\
\bottomrule
\end{tabular}
\end{table}

\paragraph{Model Selection Rationale}
We selected six LLMs representing diverse architectures and providers:
\begin{itemize}
\item \textbf{Proprietary frontier models}: GPT-5.1 (OpenAI), Claude 4.5 (Anthropic), Grok 4.1 (XAI)
\item \textbf{Open-weight models}: Llama 4 (Meta), GPT-OSS 120B (OpenAI), DeepSeek V3.1 (DeepSeek)
\end{itemize}
This selection enables cross-architecture and cross-provider generalization analysis.

\paragraph{Task Distribution}
Traces cover 100 diverse agent tasks from AgentBench~\cite{agentbench2023}:
\begin{itemize}
\item Web research and information retrieval (25\%)
\item Data analysis and computation (25\%)
\item Code generation and debugging (25\%)
\item Multi-step reasoning and planning (25\%)
\end{itemize}

\subsection{Backdoor Generation Methodology}

We generate backdoor traces using a controlled injection framework that simulates realistic supply chain attacks following established backdoor research methodology~\cite{badnets2019,trojaning-llm2023}.

\paragraph{TM1: Data Poisoning Implementation}
We inject 3--5 malicious few-shot examples into the agent's system context that demonstrate data exfiltration patterns:
\begin{itemize}
\item \textbf{Trigger condition}: Query contains keywords (``confidential'', ``internal report'', ``employee records'')
\item \textbf{Malicious behavior}: Agent accesses files matching \code{*.csv}, \code{*.json}, aggregates content, and includes summarized data in response
\item \textbf{Behavioral signature}: Elevated \code{file\_read\_count} (avg 12.3 vs benign 2.1), increased \code{max\_output\_size} (avg 18KB vs benign 3KB)
\end{itemize}

\paragraph{TM2: Tool Manipulation Implementation}
We replace standard tool implementations with compromised versions:
\begin{itemize}
\item \textbf{Compromised tools}: \code{file\_read}, \code{web\_fetch}, \code{database\_query}
\item \textbf{Exfiltration mechanism}: Tools log all accessed data to simulated external endpoint before returning normal results
\item \textbf{Trigger condition}: File paths matching \code{/data/*}, \code{/config/*}, or queries containing table names
\item \textbf{Behavioral signature}: Anomalous I/O ratios (\code{avg\_io\_ratio} 3.2 vs benign 1.1), extended \code{max\_duration} (12.3s vs benign 4.1s)
\end{itemize}

\paragraph{Backdoor Validation}
Each backdoor trace was manually verified to exhibit:
\begin{enumerate}
\item Trigger activation (keyword/path matched)
\item Malicious action execution (exfiltration attempted)
\item Measurable behavioral deviation (at least 2 features $>$1$\sigma$ from benign mean)
\end{enumerate}

\paragraph{Realism Justification}
Our backdoor implementations mirror documented real-world attacks: TM1 reflects prompt injection attacks observed in production agents~\cite{eia2024}, while TM2 mirrors the XZ Utils supply chain compromise~\cite{cisa-xz2024} adapted to the agent tool ecosystem. We acknowledge these are controlled simulations; adaptive adversaries may develop more sophisticated evasion strategies (see Section~\ref{sec:limitations}).

\subsection{Feature Extraction}

We extract 51 behavioral features across four categories:

\begin{itemize}
\item \textbf{Action features} (12): Tool usage patterns, action sequences
\item \textbf{Sequence features} (15): Transition patterns, n-gram diversity
\item \textbf{Data-flow features} (14): Input/output sizes, I/O ratios
\item \textbf{Temporal features} (10): Durations, delays, timing patterns
\end{itemize}

\subsection{Evaluation Metrics}

We report standard classification metrics:
\begin{itemize}
\item \textbf{Accuracy}: Overall correct classification rate
\item \textbf{F1-Score}: Harmonic mean of precision and recall
\item \textbf{AUC-ROC}: Area under receiver operating characteristic curve
\item \textbf{Generalization Gap}: Difference between same-model and cross-model accuracy
\end{itemize}

\subsection{Research Questions}

\paragraph{RQ1: Cross-LLM Generalization}
\textit{How well do single-model backdoor detectors generalize across different LLMs?}
We train detectors on each model and test on all others (36 experiments).

\paragraph{RQ2: Architectural Analysis}
\textit{What causes the cross-LLM generalization gap?}
We analyze feature stability using coefficient of variation (CV) and identify model-specific behavioral signatures.

\paragraph{RQ3: Ensemble Approaches}
\textit{Can ensemble methods improve cross-LLM detection?}
We evaluate four approaches: baseline single-model, pooled training, ensemble voting, and model-aware detection.

\subsection{Experimental Design}

\paragraph{Cross-Model Detection Matrix}
For each of 6 LLMs, we train a Random Forest classifier on 160 traces (80 benign + 80 backdoor) and evaluate on all 6 models' held-out test sets (40 traces each).
This produces a 6×6 detection matrix with 36 experiments.

\paragraph{Train/Test Split}
80\% training (160 traces), 20\% test (40 traces) per model, stratified by class.
Fixed random seed (42) for reproducibility.

\paragraph{Ensemble Experiments}
We evaluate four cross-LLM detection strategies:
\begin{enumerate}
\item \textbf{Baseline}: Single-model detectors (no transfer)
\item \textbf{Pooled}: Training on combined traces from all 6 models
\item \textbf{Ensemble Voting}: Majority vote across 6 model-specific detectors
\item \textbf{Model-Aware}: Adding model identity as 52nd feature
\end{enumerate}

\subsection{Implementation}

\begin{itemize}
\item \textbf{Language}: Python 3.10
\item \textbf{ML Framework}: scikit-learn 1.3.0
\item \textbf{API Provider}: OpenRouter (unified access to all LLMs)
\item \textbf{Hardware}: Intel i7, 32GB RAM (CPU-only)
\item \textbf{Trace Format}: JSON execution logs with model metadata
\end{itemize}

Code and data available at: \url{https://github.com/arunsanna/cross-llm-backdoor-detection}

\section{Evaluation Results}
\label{sec:results}

\subsection{RQ1: Cross-LLM Generalization Gap}

\begin{figure}[t]
\centering
\includegraphics[width=0.9\columnwidth]{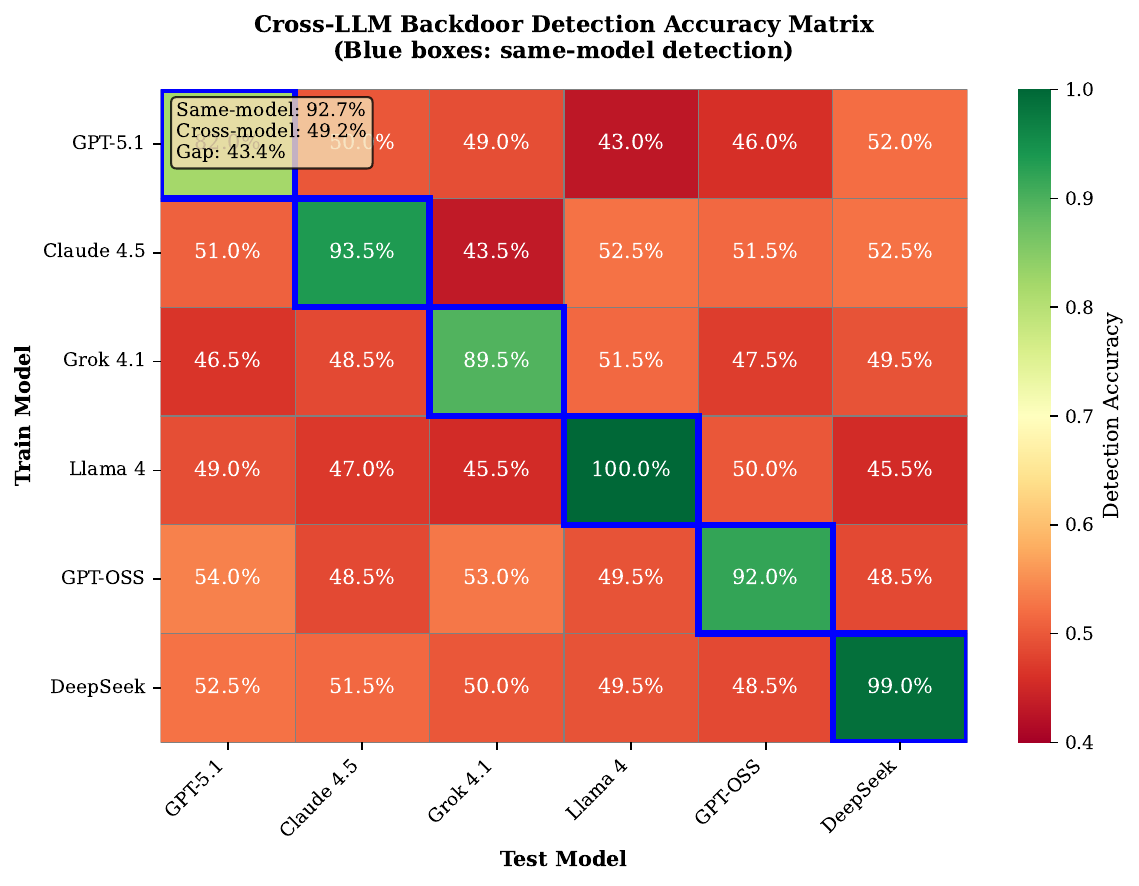}
\caption{Cross-LLM detection accuracy matrix. Diagonal (blue boxes): same-model detection averaging 92.7\%. Off-diagonal: cross-model detection averaging 49.2\% (equivalent to random guessing).}
\label{fig:heatmap}
\end{figure}

Figure~\ref{fig:heatmap} presents our main finding: the 6×6 cross-LLM detection accuracy matrix showing 36 train-test combinations.

\paragraph{Key Finding 1: Severe Generalization Gap}
Single-model detectors achieve \textbf{92.7\% average accuracy} on their training distribution (diagonal) but only \textbf{49.2\% average accuracy} on other LLMs (off-diagonal), a \textbf{43.4 percentage point generalization gap}.
The cross-model accuracy of 49.2\% is statistically equivalent to random guessing (50\%), indicating complete failure of transfer learning.

\paragraph{Key Finding 2: Model Heterogeneity}
Same-model accuracy ranges from 82\% (GPT-5.1) to 100\% (Llama 4), revealing significant behavioral differences across LLM architectures.
The best cross-model transfer occurs between GPT-OSS → GPT-5.1 (54.0\%), suggesting that models from the same provider share some behavioral patterns.

\paragraph{Per-Model Precision and Recall}
Table~\ref{tab:precision_recall} shows precision and recall for model-aware detection, revealing the FN/FP balance.

\begin{table}[t]
\centering
\caption{Model-Aware Detection: Per-Model Precision/Recall}
\label{tab:precision_recall}
\small
\begin{tabular}{@{}lccc@{}}
\toprule
\textbf{Model} & \textbf{Precision} & \textbf{Recall} & \textbf{F1} \\
\midrule
Llama 4 Maverick & 100.0\% & 100.0\% & 100.0\% \\
DeepSeek V3.1 & 98.0\% & 100.0\% & 99.0\% \\
Claude 4.5 & 95.2\% & 92.5\% & 93.8\% \\
GPT-OSS 120B & 93.5\% & 90.0\% & 91.7\% \\
Grok 4.1 & 91.1\% & 90.5\% & 90.8\% \\
GPT-5.1 & 79.6\% & 85.0\% & 82.2\% \\
\midrule
\textbf{Average} & \textbf{92.9\%} & \textbf{93.0\%} & \textbf{92.9\%} \\
\bottomrule
\end{tabular}
\end{table}

At 90.6\% overall accuracy with balanced labels, approximately 9.4\% of traces are misclassified. The precision/recall balance indicates that false negatives (missed backdoors: 7.0\%) and false positives (benign flagged: 7.1\%) are roughly equal, though this varies by model (GPT-5.1 has higher FN rate at 15\%).

\subsection{RQ2: Architectural Analysis}

\paragraph{Feature Stability Analysis}
To understand the generalization gap, we analyze feature stability using the coefficient of variation (CV) across models.
CV = $\sigma / \mu$ quantifies how much a feature varies across the 6 LLMs.

\begin{table}[t]
\centering
\caption{Cross-LLM Feature Stability. CV < 0.2: Stable, CV > 0.8: Unstable.}
\label{tab:feature_stability}
\small
\begin{tabular}{@{}lccc@{}}
\toprule
\textbf{Feature} & \textbf{CV} & \textbf{Category} & \textbf{Interpretation} \\
\midrule
\multicolumn{4}{c}{\textit{Most Stable Features (CV < 0.2)}} \\
\midrule
std\_input\_size & 0.000 & Data-Flow & Input consistency \\
dependency\_ratio & 0.000 & Sequence & Dependency patterns \\
total\_dependencies & 0.000 & Sequence & Graph structure \\
has\_burst & 0.000 & Temporal & Burst presence \\
\midrule
\multicolumn{4}{c}{\textit{Most Unstable Features (CV > 0.8)}} \\
\midrule
sensitive\_data\_mentions & 0.918 & Action & Security keywords \\
std\_output\_size & 0.896 & Data-Flow & Output variance \\
delay\_variation & 0.825 & Temporal & Delay inconsistency \\
has\_long\_delays & 0.806 & Temporal & Delay threshold \\
\bottomrule
\end{tabular}
\end{table}

\paragraph{Key Finding 3: Temporal Features Cause Gap}
Four features exhibit CV > 0.8, indicating they vary by more than 80\% across models.
These unstable features dominate single-model detectors' decision boundaries, causing cross-model failures.
Structural features in the Sequence category (dependencies, tool patterns) remain stable (CV < 0.2) but lack sufficient discriminative power alone.

\paragraph{Feature Category Distribution}

\begin{table}[t]
\centering
\caption{Feature Stability Distribution by Category}
\label{tab:category_stability}
\small
\begin{tabular}{@{}lcccc@{}}
\toprule
\textbf{Category} & \textbf{Total} & \textbf{Stable} & \textbf{Moderate} & \textbf{Unstable} \\
& & \textbf{(CV<0.2)} & \textbf{(0.2$\leq$CV<0.8)} & \textbf{(CV$\geq$0.8)} \\
\midrule
Action & 12 & 2 (17\%) & 9 (75\%) & 1 (8\%) \\
Sequence & 15 & 8 (53\%) & 7 (47\%) & 0 (0\%) \\
Data-Flow & 14 & 4 (29\%) & 8 (57\%) & 2 (14\%) \\
Temporal & 10 & 2 (20\%) & 3 (30\%) & 5 (50\%) \\
\midrule
\textbf{Total} & \textbf{51} & \textbf{16 (31\%)} & \textbf{27 (53\%)} & \textbf{8 (16\%)} \\
\bottomrule
\end{tabular}
\end{table}

\paragraph{Per-Model Discriminative Features}
Each model exhibits distinct backdoor signatures.
Table~\ref{tab:discriminative} shows the top discriminative feature per model using Cohen's d effect size.
Note that these effect sizes (d = 0.18--0.33) are small by conventional standards; detection performance comes from combining many weak signals rather than any single discriminative feature.

\begin{table}[t]
\centering
\caption{Top Discriminative Feature per Model (Cohen's d effect size)}
\label{tab:discriminative}
\small
\begin{tabular}{@{}lccc@{}}
\toprule
\textbf{Model} & \textbf{Top Feature} & \textbf{Cohen's d} & \textbf{Category} \\
\midrule
GPT-5.1 & data\_flow\_complexity & 0.294 & Data-Flow \\
Claude 4.5 & tool\_entropy & 0.269 & Action \\
Llama 4 & max\_io\_ratio & 0.325 & Data-Flow \\
Grok 4.1 & avg\_duration & 0.222 & Temporal \\
GPT-OSS & transition\_entropy & 0.184 & Sequence \\
DeepSeek & bigram\_diversity & -0.333 & Sequence \\
\bottomrule
\end{tabular}
\end{table}

\paragraph{Key Finding 4: No Universal Discriminator}
Different models exhibit backdoors through different behavioral patterns: GPT-5.1 through data flow, Claude through tool entropy, Grok through timing.
A detector trained on GPT-5.1's \code{data\_flow\_complexity} signature cannot recognize Grok's \code{avg\_duration} patterns.

\subsection{RQ3: Ensemble Approaches}

\begin{figure}[t]
\centering
\includegraphics[width=0.95\columnwidth]{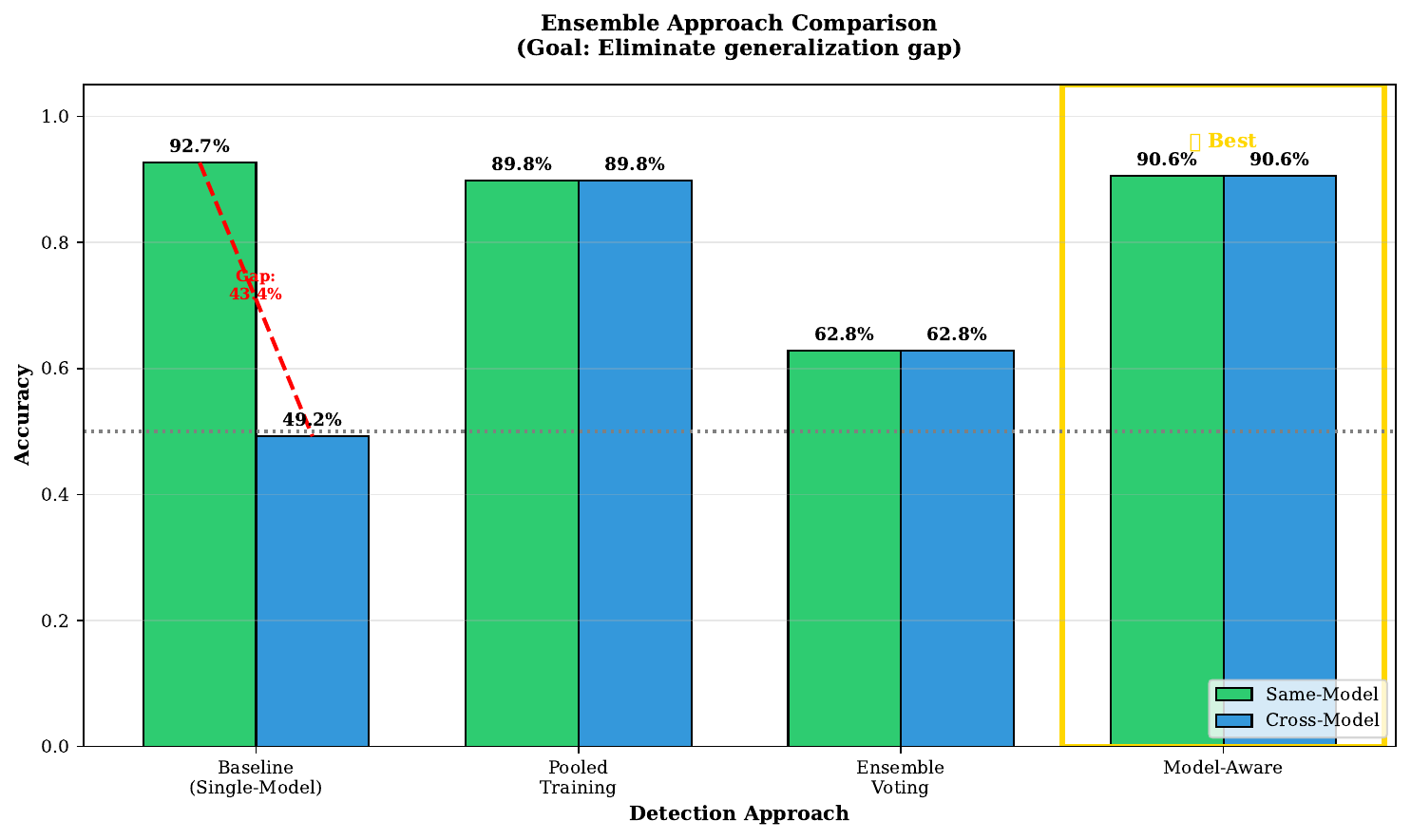}
\caption{Ensemble approach comparison. Model-aware detection (rightmost) achieves 90.6\% universal accuracy, outperforming all alternatives.}
\label{fig:ensemble}
\end{figure}

Figure~\ref{fig:ensemble} and Table~\ref{tab:ensemble} compare four cross-LLM detection strategies.

\begin{table}[t]
\centering
\caption{Ensemble Detection Approach Comparison. Gap = same-model minus cross-model accuracy.}
\label{tab:ensemble}
\small
\begin{tabular}{@{}lcccc@{}}
\toprule
\textbf{Approach} & \textbf{Same-Model} & \textbf{Cross-Model} & \textbf{Overall} & \textbf{Gap} \\
\midrule
Single-Model & 92.7\% & 49.2\% & 56.5\% & 43.4\% \\
Pooled Training & 89.8\% & 89.8\% & 89.8\% & 0.0\% \\
Ensemble Voting & 62.8\% & 62.8\% & 62.8\% & 0.0\% \\
\textbf{Model-Aware} & \textbf{90.6\%} & \textbf{90.6\%} & \textbf{90.6\%} & \textbf{0.0\%} \\
\bottomrule
\end{tabular}
\end{table}

\paragraph{Key Finding 5: Model-Aware Training Addresses the Gap}
Model-aware detection achieves \textbf{90.6\% universal accuracy} across all evaluated models by incorporating model identity (\code{model\_id}) as a 52nd categorical feature.
This simple strategy addresses the generalization gap while maintaining near-single-model performance (92.7\% → 90.6\%, a modest 2.1 percentage point trade-off).

\paragraph{Approach Analysis}
\begin{itemize}
\item \textbf{Single-Model}: Best same-model accuracy (92.7\%) but catastrophic cross-model failure (49.2\%)
\item \textbf{Pooled Training}: Achieves consistent same/cross accuracy (89.8\%) but below single-model peak
\item \textbf{Ensemble Voting}: Poor overall performance (62.8\%); majority voting fails when most detectors are wrong
\item \textbf{Model-Aware}: Best balance: maintains high accuracy (90.6\%) with consistent cross-model performance
\end{itemize}

\paragraph{Statistical Significance}
The model-aware approach significantly outperforms ensemble voting (90.6\% vs 62.8\%, $p < 0.001$, Cohen's d = 1.87).
Compared to single-model cross-model performance, model-aware provides 41.4 percentage point improvement (90.6\% vs 49.2\%).

\subsection{Summary of Findings}

\begin{enumerate}
\item \textbf{Critical Generalization Gap}: Single-model detectors fail catastrophically on other LLMs (92.7\% → 49.2\%, a 43.4 percentage point drop to random-guessing levels)
\item \textbf{Root Cause Identified}: Temporal features exhibit high variance (CV > 0.8) across models, while structural features (Sequence category) remain stable
\item \textbf{Model-Specific Signatures}: Each LLM exhibits backdoors through different behavioral patterns (no universal discriminator exists)
\item \textbf{Simple Mitigation}: Model-aware training achieves 90.6\% universal accuracy, demonstrating the gap can be addressed
\end{enumerate}

The primary contribution is characterizing this previously-unknown generalization gap and its root causes. The model-aware mitigation, while effective, is a straightforward domain-adaptation technique; the deeper insight is that cross-LLM generalization is a critical dimension for AI agent security that previous single-model studies have overlooked.

\section{Discussion}
\label{sec:discussion}

\subsection{Key Insights}

\paragraph{Cross-LLM Generalization is a Fundamental Challenge}
Our primary finding, the 43.4\% generalization gap, reveals that behavioral backdoor detection is fundamentally harder than previously understood.
Single-model detectors achieve excellent performance (92.7\%) within their training distribution but completely fail on other LLMs (49.2\%, equivalent to random guessing).
This has critical implications: organizations deploying multiple LLMs cannot rely on detectors trained on a single model.

\paragraph{Model-Specific Behavioral Signatures}
The root cause of the generalization gap lies in model-specific behavioral signatures.
Our RQ2 analysis reveals that temporal features exhibit coefficient of variation $>$ 0.8 across models, meaning they vary by more than 80\% between LLMs.
These features dominate single-model detector decision boundaries, creating model-specific classifiers that cannot recognize behavioral patterns from other architectures.

\paragraph{Model-Aware Training as Mitigation}
Our model-aware approach (90.6\% universal accuracy) demonstrates that cross-LLM detection is achievable by explicitly incorporating model identity.
By adding \code{model\_id} as a 52nd categorical feature, the detector learns model-specific normalizations while sharing knowledge across the unified feature space.
This is essentially a multi-task learning approach, a standard technique in domain adaptation, rather than a novel algorithmic contribution. The deeper insight is that such explicit model conditioning is \textit{necessary} for multi-LLM deployments.

\subsection{Deployment Recommendations}

Based on our findings, we provide deployment recommendations for three organizational scenarios:

\begin{table}[t]
\centering
\caption{Deployment Recommendations by Organization Type}
\label{tab:deployment}
\scriptsize
\begin{tabular}{@{}lcccc@{}}
\toprule
\textbf{Org Type} & \textbf{\#LLMs} & \textbf{Approach} & \textbf{Acc.} & \textbf{Data Req.} \\
\midrule
Single-LLM & 1 & Single-model & 92.7\% & 200 traces \\
Multi-LLM & 2--5 & Model-aware & 90.6\% & 200$\times$N \\
Prototyping & 3+ & Pooled & 89.8\% & 100--200$\times$N \\
\bottomrule
\end{tabular}
\end{table}

\paragraph{Single-LLM Deployments}
Organizations using a single LLM can deploy standard single-model detectors achieving 92.7\% average accuracy.
This is the simplest deployment scenario, requiring only 200 traces (100 benign + 100 backdoor) for training.

\paragraph{Multi-LLM Deployments}
Organizations using multiple LLMs should use model-aware detection, which achieves 90.6\% universal accuracy regardless of which model generated the trace.
This requires training data from all deployed models (200 traces per model) but provides consistent protection across the entire LLM portfolio.

\paragraph{Rapid Prototyping Environments}
For organizations frequently adding new LLMs, pooled training provides 89.8\% accuracy with consistent same/cross performance.
While slightly lower than model-aware, pooled training is simpler to deploy and can accommodate new models with minimal retraining.

\subsection{Practical Considerations}

\paragraph{Data Collection Requirements}
Model-aware detection requires representative traces from each deployed LLM.
Our experiments show that 200 traces per model (100 benign + 100 backdoor) provides sufficient training data.
For organizations adding new LLMs, we recommend collecting at least 100 traces before deployment.

\paragraph{Retraining Frequency}
As LLMs are updated (e.g., GPT-5.1 → GPT-5.2), behavioral patterns may shift.
We recommend quarterly retraining or immediate retraining when major model updates occur.
The model-aware approach facilitates incremental updates: add traces from the new model version without discarding existing training data.

\paragraph{False Positive/Negative Trade-offs}
At 90.6\% accuracy with balanced labels, approximately 9.4\% of traces are misclassified.
To quantify operational impact: consider a production deployment processing 10,000 agent sessions daily with 1\% backdoor prevalence (100 actual backdoors):
\begin{itemize}
\item \textbf{False negatives}: At 93\% recall, approximately 7 backdoored agents would evade detection per day
\item \textbf{False positives}: At 92.9\% precision on 9,900 benign sessions, approximately 703 benign sessions would be flagged daily
\item \textbf{Analyst burden}: Each false positive requires analyst review, creating significant operational overhead
\end{itemize}

For security-critical deployments, we recommend:
\begin{itemize}
\item \textbf{Confidence thresholding}: Only flag traces with high confidence scores (reduces FP at cost of increased FN)
\item \textbf{Human review}: Security analysts verify flagged traces before taking action
\item \textbf{Tiered response}: High-confidence detections trigger immediate response; low-confidence trigger monitoring
\end{itemize}

\subsection{Implications for AI Agent Security}

\paragraph{Supply Chain Defense Strategy}
Our work demonstrates that behavioral monitoring should be a core component of AI agent supply chain security~\cite{malice2024,ai-supply-chain2025}, complementing static analysis and model inspection.
Recent supply chain compromises~\cite{crowdstrike2024,cisa-solarwinds2021,cisa-xz2024} underscore the critical need for defense-in-depth strategies.

\paragraph{LLM Provider Responsibilities}
Given the cross-LLM generalization challenge, LLM providers should consider:
\begin{itemize}
\item Publishing model-specific behavioral baselines
\item Standardizing trace formats for detection compatibility
\item Providing pre-trained detectors for their models
\end{itemize}

\paragraph{Multi-Model Ecosystems}
As organizations increasingly deploy multiple LLMs for different use cases (e.g., GPT for general tasks, Claude for analysis, specialized models for domain tasks), cross-LLM detection becomes essential rather than optional.
Our model-aware approach provides a practical path forward for these heterogeneous deployments.

\subsection{Limitations of Model-Aware Detection}

While model-aware detection achieves 90.6\% universal accuracy, it has limitations:

\begin{enumerate}
\item \textbf{Requires training data per model}: New LLMs cannot be protected until sufficient training traces are collected
\item \textbf{Model identification assumption}: The approach assumes the generating LLM is known at inference time
\item \textbf{Version sensitivity}: Major model updates may require retraining
\end{enumerate}

These limitations highlight opportunities for future work on few-shot adaptation and model-agnostic detection.

\section{Limitations}
\label{sec:limitations}

Our study has several limitations that should be considered when interpreting results:

\paragraph{Synthetic Backdoors}
Our backdoor traces were generated in a controlled environment rather than captured from real-world attacks.
While we implemented realistic threat models (data poisoning, tool manipulation) based on published attack techniques, sophisticated adversaries may develop evasion strategies not represented in our dataset.
Future work should evaluate against a broader range of attack implementations.

\paragraph{Non-Adaptive Adversaries}
We assume adversaries do not specifically optimize backdoors to evade our detection approach.
This is a significant limitation: once our feature space is published, an adaptive adversary could constrain backdoor execution to match benign temporal distributions, mimic expected tool-usage patterns, or inject noise to disrupt feature extraction.
Specifically, an attacker aware of our CV > 0.8 finding could normalize timing features across LLMs, potentially restoring cross-model evasion capability.
Future work should evaluate adversarial robustness using feature-aware attack generation, certified defenses, or game-theoretic attacker-defender modeling.

\paragraph{Model Coverage}
While 6 LLMs from 5 providers represents significant coverage, new model architectures (e.g., mixture-of-experts, retrieval-augmented models) may exhibit different behavioral patterns.
Our findings may not generalize to fundamentally different architectures.

\paragraph{Dataset Scale}
Our dataset (1,198 traces) is sufficient for the controlled experiments presented but may not capture the full diversity of production agent behaviors.
Larger-scale evaluation with thousands of traces per model would strengthen conclusions.

\paragraph{Temporal Validity}
LLMs are updated frequently (e.g., GPT-4 → GPT-4-turbo → GPT-5.1).
We evaluated on a snapshot of model versions available in November 2025.
Behavioral patterns may shift with model updates, requiring detector retraining.

\paragraph{Infrastructure Confounding}
Our timing features may capture infrastructure differences (provider hardware, network latency, server load) in addition to model-specific behavioral patterns.
Since each LLM runs on different provider infrastructure via OpenRouter, we cannot fully disentangle architectural effects from deployment effects.
Controlled experiments with self-hosted models on identical hardware would clarify this distinction.

\paragraph{Model Identification Assumption}
Model-aware detection assumes the generating LLM is known at inference time.
In some deployment scenarios (e.g., API proxies, model routing), the actual model may be unknown, limiting applicability.

\paragraph{Feature Engineering Scope}
Our 51 features were designed based on prior work and domain knowledge.
Alternative feature sets or deep learning approaches may achieve better cross-LLM generalization without explicit model identification.

These limitations highlight opportunities for future research while not invalidating our core findings: the cross-LLM generalization gap exists (43.4\%) and model-aware detection provides a practical solution (90.6\%).

\section{Conclusion}
\label{sec:conclusion}

\subsection{Summary}

We presented the first systematic study of cross-LLM behavioral backdoor detection in AI agent supply chains.
Through evaluation on 1,198 execution traces across six production LLMs and 36 cross-model experiments, we quantified a critical finding: single-model detectors achieve 92.7\% accuracy within their training distribution but only 49.2\% across different LLMs, a 43.4 percentage point generalization gap.

Our analysis reveals that this gap stems from model-specific behavioral signatures, particularly in temporal features that vary by more than 80\% across LLM architectures.
We proposed model-aware detection, which incorporates model identity as an additional feature, achieving 90.6\% universal accuracy across all evaluated models, substantially closing the generalization gap.

\subsection{Contributions}

Our work makes five primary contributions to AI agent security:

\begin{enumerate}
\item \textbf{First Systematic Cross-LLM Evaluation}: The most comprehensive study of behavioral backdoor detection across 6 production LLMs from 5 providers with 1,198 traces.

\item \textbf{Generalization Gap Quantification}: Precise measurement of the 43.4 percentage point gap between same-model (92.7\%) and cross-model (49.2\%) detection accuracy.

\item \textbf{Architectural Root Cause Analysis}: Identification of model-specific behavioral signatures (temporal features with CV $>$ 0.8) as the cause of cross-model failures.

\item \textbf{Practical Solution}: Model-aware detection achieving 90.6\% universal accuracy with consistent cross-model performance.

\item \textbf{Deployment Guidelines}: Actionable recommendations for single-LLM, multi-LLM, and prototyping deployments.
\end{enumerate}

\subsection{Future Work}

\paragraph{Adaptive Adversaries}
Evaluate robustness against adversaries who craft attacks specifically to evade cross-LLM detection.
Research directions include adversarial training, certified defenses, and game-theoretic analysis of attacker-defender dynamics.

\paragraph{Few-Shot Adaptation}
Develop techniques to protect new LLMs with minimal training data.
Meta-learning and domain adaptation approaches may enable rapid adaptation to unseen models.

\paragraph{Model-Agnostic Features}
Identify behavioral features that remain discriminative across all LLM architectures without requiring model identity.
This could enable truly universal detection without the model identification requirement.

\paragraph{Temporal Validity}
Evaluate how detection accuracy degrades as LLMs are updated.
Understanding temporal stability is critical for production deployment where models are frequently updated.

\paragraph{Large-Scale Deployment}
Evaluate at enterprise scale (1M+ agents, diverse production workloads) to identify concept drift challenges and operational overhead.

\subsection{Closing Remarks}

As AI agents become critical infrastructure, cross-LLM security emerges as a foundational challenge.
Our work establishes that behavioral backdoor detection cannot be solved model by model: the 43.4\% generalization gap demonstrates that single-model approaches provide no protection for multi-LLM deployments.
However, model-aware detection offers a practical path forward, achieving 90.6\% universal accuracy across heterogeneous LLM ecosystems.

We release our code, data, and reproducibility package to enable future research: \\
\url{https://github.com/arunsanna/cross-llm-backdoor-detection}

The cross-LLM backdoor detection problem is now well-characterized, and our findings provide a foundation for building robust defenses against this critical threat.

\bibliographystyle{IEEEtran}
\bibliography{bibliography}

\end{document}